# Evaluation of soil thermal conductivity schemes for use in land surface modelling

Yongjiu Dai[*], Nan Wei[*], Hua Yuan, Shupeng Zhang, Wei Shangguan, Shaofeng Liu, and Xingjie Lu

Southern Marine Science and Engineering Guangdong Laboratory (Zhuhai), Guangdong Province Key Laboratory for Climate Change and Natural Disaster Studies, School of Atmospheric Sciences, Sun Yat-sen University, Guangzhou 510275, China

Correspondence to:

Yongjiu Dai (daiyj6@mail.sysu.edu.cn) and Nan Wei (wein6@mail.sysu.edu.cn)

School of Atmospheric Sciences, Sun Yat-sen University

No. 135, Xingangxi Road, Guangzhou 510275, China

**Key points:**

- Seven highly recommended soil thermal conductivity schemes are evaluated by incorporation into the Common Land Model.
- The *Balland and Arp* [2005] scheme can be recommended as a superior scheme to the others for land modelling use.
- In-situ and global simulations both show a strong dependence of land surface modelling on soil thermal conductivity formulation.

**Abstract**

Soil thermal conductivity is an important physical parameter in modeling land surface processes. Previous studies on evaluations of parameterization schemes of soil thermal conductivity are mostly based on specific experimental conditions or local soil samples, and their recommendations may not be the optimal schemes for land surface model (LSMs). In this work, seven highly recommended soil thermal conductivity schemes are evaluated for their applicability in LSMs. With the consideration of both scheme estimations and land process simulations by incorporation into the Common Land Model, the *Balland and Arp* [2005] scheme is found to consistently perform best among all the schemes, and thus can be recommended as a superior scheme for land modeling use. Uncertainty analyses by in-situ simulations demonstrate that, over relatively dry regions, the inter-scheme variations of soil thermal conductivity can lead to significant differences of simulated soil temperature, especially at deep layers, due to changes of downward soil heat conduction and the associated freeze-thaw cycles. However, few effects appear over wet regions, likely due to the high soil heat capacity induced by high soil moisture levels, which increases the heat inertia in soil thermodynamics. Global comparisons show the similar relationships that soil thermal conductivity significantly affects the simulated soil temperature and other related thermal and hydraulic variables over arid and semi-arid regions in mid- and high-latitudes. These results display the role of soil thermal conductivity in LSM, and suggest the importance of the evaluation and further development of thermal conductivity schemes with respect to land modelling applications.

## 1. Introduction

Soil thermal properties (generally refer to heat capacity and thermal conductivity) are greatly important in land surface processes modelling as they influence a wide range of physical, biological and chemical processes through regulating energy partitioning at the ground surface and energy distribution at subsurface soil layers [e.g. *Peters-Lidard et al.*, 1998; *Ochsner et al.*, 2001; *Lawrence and Slater*, 2008; *Luo et al.*, 2009; *Pan et al.*, 2017; *Zhao et al.*, 2018]. The soil thermal conductivity, which quantifies the rate of heat transfer across different soil layers, directly determines soil heat fluxes and soil temperature profiles. With phase change of soil water occurring, the soil thermal conductivity also affects soil freeze-thaw cycles and soil water movement [*Cuntz and Haverd*, 2018; *Wang and Yang*, 2018]. Given the strong control on soil thermodynamics, the soil thermal conductivity is always considered as one of the most important physical parameters in land modelling studies [*Peters-Lidard et al.*, 1998; *Luo et al.*, 2009; *Zhao et al.*, 2018].

Accurate estimation of soil thermal conductivity is always difficult, since it requires too much soil information such as mineral composition (particularly quartz content), particle size distribution, porosity, dry density, soil moisture and soil temperature [e.g. *Farouki*, 1981; *Peters-Lidard et al.*, 1998]. With these controlling factors, many theoretical and empirical models have been developed to predict soil thermal conductivity [*Farouki*, 1981; *Dong et al.*, 2015]. For example, *Wiener* [1912] found a theoretical upper and lower limit for the prediction of soil thermal conductivity, which were provided by the parallel and series flow models, respectively. *De Vries* [1963] developed a Maxwell equation analogous model which gave a theoretical description of soil thermal conductivity as the weighted average of thermal conductivity from each soil constituent. *Johansen* [1975] provided a classical empirical method to estimate soil thermal conductivity by interpolating between the values of dry and saturated soil, where the weighting coefficient of saturated soil (also called relative thermal conductivity or the Kersten number) was obtained by fitting a logarithmic function with the degree of saturation using experimental measurements. Johansen's method also brought up some other empirical models, in which new relationships between relative thermal conductivity and degree of saturation, porosity or soil types were proposed [e.g. *Côté and Konrad*, 2005; *Lu et al.*, 2007]. A recent comprehensive review of soil thermal conductivity models for unsaturated soils can be found in *Dong et al.* [2015]. They analyzed key factors affecting soil thermal conductivity, and pointed out the common pitfalls among all the existing models.

For soil thermal conductivity models or schemes, a lot of work has been conducted for their comparisons and evaluations. *Farouki* [1981] compared 11 early schemes including the *De Vries* [1963] model and the *Johansen* [1975] model, and found that the *Johansen* [1975] model gave the most accurate prediction to thermal conductivity by comparison with experimental measurements over a range of soil types and saturation levels. *Barry-Macaulay et al.* [2015] compared the *Johansen* [1975] model with its three derivative models developed by *Balland and Arp* [2005], *Côté and Konrad* [2005] and *Lu et al.* [2007], and they showed that all the four models gave good fit to experimental data, with the best agreement provided by the *Côté and Konrad* [2005] model. *Zhang and Wang* [2017] presented a thorough summary for advantages and disadvantages of 13 typical schemes, and they demonstrated that the methods of *Chen* [2008], *Haigh* [2012] and *Zhang* et al. [2015] were superior to other schemes in predicting sand thermal conductivity. Although these works and many other related studies try to tell which thermal conductivity scheme is the best, most of them are based on specific experimental conditions or local soil samples, and few studies can provide effective evidence to show which scheme is more reliable for land modelling applications. This suggests the need to couple soil thermal conductivity schemes with land surface models to reevaluate these schemes for regional or global land modelling use.

Based on early soil property datasets, most thermal conductivity schemes only focus on information from fine mineral soils (particle diameter < 2 mm), while few have considered the effects of soil organic matter (SOM) or gravels (particle diameter > 2 mm). SOM and gravels have been proved to have different hydraulic and thermal properties (e.g. relatively low thermal conductivity and high heat capacity for SOM and high thermal conductivity for gravels) than fine mineral soils, and thus can significantly affect soil moisture and temperature simulations and even alter dynamics of boundary layer development in global climate models [e.g. *Lawrence and Slater*, 2008; *Pan et al.*, 2017]. With the enrichment of soil information collected, fractions of SOM and gravels have been gradually complied into several soil datasets such as Global Soil Dataset for Earth system model (GSDE) [*Shangguan et al.*, 2014] and SoilGrids [*Hengl et al.*, 2017]. These datasets make it possible to add the effects of SOM and gravels into existing soil thermal conductivity schemes and thus a more complete description can be implemented in land surface models.

The presented paper focuses mostly on applications of soil thermal conductivity schemes in land surface processes modelling. Seven highly recommended thermal conductivity schemes are selected with the modification of containing SOM and gravel effects. Besides direct comparison with laboratory measured thermal conductivities, these schemes are incorporated into the Common Land Model (CoLM), which is the land component in many earth system models (e.g. CWRF [*Liang et al.*, 2012] and BNU-ESM [*Ji et al.*, 2014]), so that they can be reevaluated based on their corresponding CoLM performances by comparison with in-situ observed land state variables such as soil temperature and ground heat flux. Moreover, the uncertainties of land surface modelling introduced by thermal conductivity formulation can also be quantified through analyzing the discrepancies of CoLM performances. This work is expected to identify a superior thermal conductivity scheme for use with consideration of the accuracy of both the scheme estimation and land process simulation, and clarify the importance of soil thermal conductivity parameterization on land surface modelling. The results could contribute to fill cognitive gaps between physical soil scientists and land process modelers.

## 2. Materials and methodology

### 2.1 *Soil thermal conductivity schemes*

In this work, seven highly recommended soil thermal conductivity schemes are selected for analysis. Five of them are from *Johansen* [1975] and its derivatives (*Farouki* [1981], *Côté and Konrad* [2005], *Balland and Arp* [2005] and *Lu et al.* [2007]), and the other two are an empirical scheme from *Tarnawski and Leong* [2012] and a theoretical scheme from *De Vries* [1963], respectively. Rather than following their original formulation which are mostly based on mineral soil fractions, all the schemes are modified as a form of volumetric weighted combination of mineral soils, SOM and gravels in order to accommodate all the effects of available soil components. The key formulation of each scheme implemented in CoLM are described below, and more details can be found in *Dai et al.* [2014].

### 2.1.1 *Johansen* [1975] scheme

*Johansen* [1975] calculates soil thermal conductivity $k$ (W/m/K) as a combination of dry $k_{dry}$ and saturated $k_{sat}$ thermal conductivities, weighted by the Kersten number $K_e$ as Equation 1 depicts:

$$k = (k_{sat} - k_{dry})K_e + k_{dry} \tag{1}$$

where the weight coefficient $K_e$ is expressed as a function of the degree of saturation [$S_r = \theta/\theta_s$, $\theta$, and $\theta_s$ are real and saturated soil moisture (cm$^3$/cm$^3$), respectively], the phase of water and soil particle size:

$$K_e = \begin{cases} 0.7\log S_r + 1, & \text{for unfrozen coarse-grained soils with } S_r > 0.05 \\ \log S_r + 1, & \text{for unfrozen fine-grained soils with } S_r > 0.1 \\ S_r, & \text{for frozen medium and fine sands and fine-grained soils} \end{cases} \tag{2}$$

The dry thermal conductivity $k_{dry}$ is estimated using the weighted arithmetic mean of thermal conductivities of mineral soils, SOM and gravels:

$$k_{dry} = f_{minerals}k_{minerals\_dry} + f_{om}k_{om\_dry} + f_{gravel}k_{gravel} \tag{3}$$

where $f_{minerals}$, $f_{minerals}$, $f_{om}$, and $f_{gravel}$ are volumetric fractions of mineral soils, SOM

and gravels in soil solids, and $k_{\text{minerals\_dry}}$, $k_{\text{om\_dry}}$, and $k_{\text{gravel}}$ are corresponding thermal conductivities in dry conditions, empirically valued as:

$$k_{\text{minerals\_dry}} = \frac{0.135\rho_d + 0.0647}{2.7 - 0.947\rho_d} \tag{4}$$

$$k_{\text{om\_dry}} = 0.05 \tag{5}$$

$$k_{\text{gravel}} = 0.039\theta_s^{-2.2} \tag{6}$$

where $\rho_d$ is dry bulk density of mineral soils (g/cm³).

The saturated thermal conductivity $k_{\text{sat}}$ is approximated as the weighted geometric mean of thermal conductivities of soil solids and water:

$$k_{\text{sat}} = k_{\text{minerals\_wet}}^{v_m} k_{\text{om\_wet}}^{v_{om}} k_{\text{gravel}}^{v_g} k_w^{\theta_s} \tag{7}$$

where $v_m$, $v_{om}$, and $v_g$ are volumetric fractions of mineral soils, SOM and gravels in all soil constituents (including water), and $k_{\text{minerals\_wet}}$, $k_{\text{om\_wet}}$ and $k_{\text{gravel}}$ are corresponding thermal conductivities in wet conditions. Here, $k_{\text{minerals\_wet}}$ depends on quartz content:

$$k_{\text{minerals\_wet}} = k_q^{v_q} k_o^{1-v_q} \tag{8}$$

where $v_q$ is volumetric fractions of quartz in mineral soils, and $k_q$ (8.8 W/m/K) and $k_o$ (2.0 W/m/K for $v_q > 0.2$, and 3.0 W/m/K for otherwise) are thermal conductivities of quartz and non-quartz minerals. $k_{\text{om\_wet}}$ is given as 0.25 W/m/K and $k_{\text{gravel}}$ uses the same estimation as that in dry conditions. $k_w$ is the thermal conductivity of water, and valued as 2.29 W/m/K for frozen and 0.57 W/m/K for unfrozen status.

*Johansen* [1975] first proposes a method to depict thermal conductivity of dry soil, and what's more important, first introduces Kersten number $K_e$ to calculate bulk thermal conductivity, where the $K_e$ - $S_r$ relationship contains the effects of soil components, porosity and water content simultaneously. Based on the *Johansen* [1975] scheme, a series of derivative schemes have been developed afterwards, since the effects of soil properties on thermal conductivity can be studied through different $K_e$-$S_r$ relationship for any given soil types. The following four derivatives of *Johansen* [1975] are all based on Equation 1, with only discrepancies in $K_e$-$S_r$ relationship as well as in formulations of thermal conductivities of dry or saturated soils.

### 2.1.2 *Farouki* [1981] scheme

*Farouki* [1981] uses the same equation proposed by *Johansen* [1975] to predict thermal conductivity of dry soil (Equation 3-6). In wet conditions, however, instead of using quartz content in calculation of thermal conductivity of mineral soils, *Farouki* [1981] directly uses the content of sand and clay, and the bulk thermal conductivity of soil solids is estimated as the weighted arithmetic mean of thermal conductivities of each soil component. Thus, the saturated thermal conductivity $k_{sat}$ is empirically calculated as:

$$k_{sat} = \left( f_{minerals} k_{minerals\_wet} + f_{om} k_{om\_wet} + f_{gravel} k_{gravel} \right)^{1-\theta_s} k_w^{\theta_s} \tag{9}$$

$$k_{minerals\_wet} = \frac{8.80\%sand + 2.92\%clay}{\%sand + \%clay} \tag{10}$$

where %sand and %clay represent the gravimetric fractions of sand and clay in mineral soils, respectively. *Farouki* [1981] also simplifies the $K_e$-$S_r$ relationship by ignoring the effects of soil particle size:

$$K_e = \begin{cases} \log S_r + 1, & \text{for unfrozen soils} \\ S_r, & \text{for frozen soils} \end{cases} \tag{11}$$

The *Farouki* [1981] scheme is easily implemented due to its simplification, and the required inputs such as the content of sand, clay, SOM and gravel can be directly obtained from soil property datasets. For these reasons, the *Farouki* [1981] scheme is always adopted as the default scheme to calculate soil thermal conductivity in land surface models such as CoLM and the Community Land Model (CLM).

### 2.1.3 *Côté and Konrad* [2005] scheme

*Côté and Konrad* [2005] analyze a large dataset of measured thermal conductivity of dry soils, and they demonstrate that the relationship between dry thermal conductivity and porosity depends strongly on soil particle shape and grain size, which is not reflected in the *Johansen* [1975] scheme. Moreover, the logarithmical $K_e$-$S_r$ relationship proposed by *Johansen* [1975] can lead to negative predictions of thermal conductivities in dry conditions, which are unrealistic values for numerical applications. Due to these limitations, *Côté and Konrad* [2005] propose a new method to predict dry thermal conductivity, and also build a new $K_e$-$S_r$ relationship to account for effects of soil wetness on thermal conductivity estimations. The thermal conductivity for saturated soils is calculated in the same way as *Johansen* [1975] did (Equation 7-8).

The new method to calculate dry thermal conductivity $k_{dry}$ is formulated as follows:

$$k_{dry} = \sum_i f_i \chi_i \left(10^{-\eta_i \theta_s}\right), \quad i = \text{minerals, SOM and gravel} \qquad (12)$$

where $\chi_i$ (W/m/K) and $\eta_i$ are empirical soil type parameters that account for particle shape effects of mineral soils, SOM and gravels. The new $K_e$-$S_r$ relationship is proposed incorporating a parameter $\kappa$:

$$K_e = \frac{\kappa S_r}{1 + (\kappa - 1) S_r} \qquad (13)$$

where $\kappa$ is an empirical parameter that is a function of soil type and phase of water. The suggested values of $\chi_i$, $\eta_i$ and $\kappa$ are given in *Côté and Konrad* [2005]. *Barry-Macaulay et al.* [2015] have shown that the *Côté and Konrad* [2005] scheme can provide the best fit to experimental measurements among several derivative schemes of *Johansen* [1975].

2.1.4 *Balland and Arp* [2005] scheme

*Balland and Arp* [2005] follow the original formulation of *Johansen* [1975] to calculate dry and saturated thermal conductivities (Equation 3-8), but they point out a fundamental flaw of the *Johansen* [1975] scheme that it fails to be seamless in estimating soil thermal conductivity from dry to saturated and from fine- to coarse-textured soils. To solve this issue, they introduce a new function to build $K_e - S_r$ relationship:

$$K_e = \begin{cases} S_r^{0.5(1+v_{om}-\alpha v_s - v_g)} \left\{ \left[\dfrac{1}{1+e^{-\beta S_r}}\right]^3 - \left(\dfrac{1-S_r}{2}\right)^3 \right\}^{1-v_{om}}, & \text{for unfrozen soils} \\ S_r^{1+v_{om}}, & \text{for frozen soils} \end{cases} \quad (14)$$

where $v_s$, $v_{om}$, and $v_g$ are volumetric fractions of sand, SOM and gravels in all soil constituents, and $\alpha\,(0.24 \pm 0.04)$ and $\beta\,(18.1 \pm 1.1)$ are adjustable parameters which are determined based on unfrozen experimental data by *Kersten* [1949] and *Ochsner et al.* [2001]. This function results in a continuous variation of predicted thermal conductivity over the entire range of soil saturation and particle size, and with this function, the excellent performance of the *Balland and Arp* [2005] scheme has been corroborated by verification with various experimental datasets [*Zhang and Wang*, 2017].

2.1.5 *Lu et al.* [2007] scheme

*Lu et al.* [2007] conduct laboratory measurements for thermal conductivities of twelve different soils over multiple saturation levels, and show that the *Johansen* [1975] scheme cannot always account for situations at low saturations, especially for fine-grained soils. They also validate the *Côté and Konrad* [2005] scheme against these measurements, and find that the $K_e$-$S_r$ relationship is extremely sensitive to the parameter $\kappa$, which therefore introduces too large uncertainties for thermal conductivity predictions. To improve the $K_e$-$S_r$ relationship, *Lu et al.* [2007] propose the following function for use:

$$K_e = \exp\left\{\alpha\left[1 - S_r^{(\alpha-\beta)}\right]\right\} \quad (15)$$

where $\alpha$ and $\beta$ are soil texture (fine or coarse) dependent variables (see their fitted values in *Lu et al.* [2007]). In addition, when calculating dry and saturated thermal conductivities as *Johansen* [1975] did, a linear relationship for dry mineral soils with porosity is used by fitting measurement samples:

$$k_{minerals\_dry} = 0.56\theta_s + 0.51 \quad (16)$$

The *Lu et al.* [2007] scheme has been shown the best fit to fine-grained soils in previous evaluation studies, but it is relatively less accurate for coarse-grained soils at low saturations [*Barry-Macaulay et al., 2015; Zhang and Wang, 2017*].

2.1.6 *Tarnawski and Leong* [2012] scheme

In the *Johansen* [1975] scheme and its derivatives mentioned above, soil components in dry conditions are assumed to be parallel to heat flow, in which the bulk thermal conductivity is given by the arithmetic mean, while in saturated conditions, the bulk thermal conductivity is calculated as the geometric mean, indicating that the soil solids and water are arranged in a way between parallel and series to heat flow. As a more complex scheme, *Tarnawski and Leong* [2012] propose a mixed series-parallel arrangement for soil constituents. They assume that heat is conducted through three pathways, namely a solid uniform passage ($\Theta_{sb}$), a series-parallel passage composed of solids connected to a parallel path of a portion of soil water ($n_w$) and a portion of soil air ($n_a$), and a path of water ($\Theta_w$) and air ($\Theta_a$) in a parallel arrangement. The symbols in the parentheses represent the volumetric fractions of each portion in a soil.

By applying a classical resistor model to each heat pathway, the bulk thermal conductivity can be obtained by the following expression:

$$k = k_s \Theta_{sb} + \frac{(1-\theta_s-\Theta_{sb}+n_{wm})^2}{\dfrac{1-\theta_s-\Theta_{sb}}{k_s} + \dfrac{n_{wm}}{k_w \dfrac{n_w}{n_{wm}} + k_a(1-\dfrac{n_w}{n_{wm}})}} + k_w \left( \theta_s S_r - n_{wm} \dfrac{n_w}{n_{wm}} \right)$$
$$+ k_a \left[ \theta_s (1-S_r) - n_{wm} \left( 1 - \dfrac{n_w}{n_{wm}} \right) \right] \tag{17}$$

where $k_s (= k_{\min erals\_wet}^{v_m} k_{om\_wet}^{v_{om}} k_{gravel}^{v_g})$, $k_w$ and $k_a$ are thermal conductivities of soil solids, water and air. $\Theta_{sb}$ and $n_{wm}$ can be empirically calculated by fitting to content of gravel and sand:

$$\Theta_{sb} = 0.0237 - 0.0175(m_{gravel} + m_{sand})^3 \tag{18}$$

$$n_{wm} = n_a + n_w = 0.088 - 0.037(m_{gravel} + m_{sand})^3 \tag{19}$$

where $m_{gravel}$ and $m_{sand}$ are gravimetric fractions of gravel and sand. The portion of soil water ($n_w$) in the series-parallel passage can be obtained by the following relationship:

$$\frac{n_w}{n_{wm}} = \begin{cases} 0, & S_r = 0 \\ \exp(1 - S_r^{-X}), & 0 < S_r \leq 1 \end{cases} \quad (20)$$

where $X = 0.6 - 0.3(m_{gravel} + m_{sand})^3$ is a minuscule pore water retention factor.

The *Tarnawski and Leong* [2012] scheme has been verified through multiple soil samples, and shows excellent performance among similar schemes especially at high saturations [*Tarnawski et al.*, 2018].

2.1.7 *De Vries* [1963] scheme

The *De Vries* [1963] scheme is a classical theoretical scheme which is developed from the Maxwell equation. In this scheme, the soil structure is assumed to be composed of ellipsoidal grains freely floating in a continuous pore fluid (air or water). Thus, the bulk thermal conductivity is estimated as a weighted average of soil constituents with their shape factors considered:

$$k = \frac{F_w \theta_s S_r k_w + F_a \theta_s (1 - S_r) k_a + F_s (1 - \theta_s) k_s}{F_w \theta_s S_r + F_a \theta_s (1 - S_r) + F_s (1 - \theta_s)} \quad (21)$$

where $F_s$, $F_w$, and $F_a$ are shape factors of soil solids, water and air. These factors are empirically given as:

$$F_i = \frac{1}{3}\left[\frac{2}{1 + (k_i/k_w - 1)g_i} + \frac{1}{1 + (k_i/k_w - 1)(1 - 2g_i)}\right], \quad i = w, a, s \quad (22)$$

where $g_i$ is a fitting parameter for ellipsoidal particles, valued depending on different soil constituents as follows:

$$g_i = \begin{cases} 0.013 + 0.944\theta_s S_r, & \text{where } \theta_s S_r \leq 0.09 \text{ for } i = a \\ 0.333 - (0.333 - 0.035)(1 - S_r), & \text{otherwise for } i = a \\ 0.125, & \text{for } i = s \end{cases} \quad (23)$$

The *De Vries* [1963] scheme can accurately predict soil thermal conductivity for most situations, but previous studies have suggested a correction coefficient of 1.0-1.25 for dry soils as this scheme always underestimates thermal conductivity under dry conditions [*Zhang and Wang,* 2017].

*2.2 Methodology*

2.2.1 Scheme evaluation with respect to laboratory measurements

The seven soil thermal conductivity schemes are first evaluated through a direct comparison of their predictions with laboratory measured values to examine the accuracy of each scheme. The measurements are from 40 Canadian soils and 21 other soils from Italy, China and Japan, which cover a wide diversity of soil conditions from loose to compact, organic to mineral, fine to coarse textured, and dry to wet [*Tarnawski et al.,* 2015; 2018]. The Canadian soils provide measurements at a full range of saturation levels ($S_r$ = 0, 0.1, 0.25, 0.5, 0.7, 1), and the other soils are measured in dryness ($S_r$ = 0) and full saturation ($S_r$ = 1). The measured thermal conductivities of Canadian soils as a function of saturation levels and porosity are shown in Figure 1. The soils consistently display an increasing trend of thermal conductivity with increasing saturation level, especially for those compacted with small porosity. This is expected as the thermal conductivity of water is far much greater than air, and thus the formed water films and bridges around soil particles can significantly reduce thermal resistance between particle surface and pore fluid and promote inter-particle heat conduction. Also, the thermal conductivities of these soils approximately increase linearly as the porosity decreases, since soil solids always have greater thermal conductivities than water and air. Note that these thermal conductivities vary in a larger range with porosity when the soils stay in higher saturations, which suggests that soil water can highlight the effects of porosity on soil heat conduction, and the discrepancies of thermal conductivities among soils are thus amplified. Similar patterns of measurements can be obtained from the 21 other soils as well. These patterns provide rationality to use these measurements as a reference to verify the soil thermal conductivity schemes. More physical details about the selected soil samples are given by *Tarnawski et al.* [2015; 2018].

2.2.2 Scheme evaluation with respect to CoLM performances

In order to evaluate the soil thermal conductivity schemes in land surface modelling, the CoLM performances incorporating different schemes are inter-compared. The CoLM is a community effort which is primarily developed and maintained by Chinese researchers. The initial version was adopted as the CLM2.0 for use with the version 2 of the Community Climate System Model [*Bonan et al.*, 2002]. Afterwards, it underwent further development in China in many areas, such as the two-big-leaf model for calculating leaf temperatures and photosynthesis-stomatal resistance and the two-stream approximation model for simulating canopy radiation [*Dai et al.*, 2004]. To date, two versions of CoLM have been released: CoLM2005 [*Dai et al.*, 2003] and CoLM2014 [*Dai et al.*, 2014]. The updates of the latter version concentrate mostly on global land surface data, pedotransfer functions for soil hydraulic and thermal parameters, and the numerical solution of the Richards equation for simulating soil water movement. *Li et al.* [2017] have compared the two versions of CoLM, and demonstrated that CoLM2014 outperformed CoLM2005 in many aspects of energy and water budget simulations. Thus, the version of CoLM2014 is selected for our land modeling runs and analyses. In CoLM, the soil heat conduction is solved numerically via the diffusion equation:

$$c\frac{\partial T}{\partial t} = \frac{\partial}{\partial z}\left[k\frac{\partial T}{\partial z}\right] \quad (24)$$

where $c$ is volumetric soil heat capacity (J/m$^3$/K), $k$ is soil thermal conductivity (W/m/K), $T$ is temperature (K), $t$ is time (s), and $z$ is soil depth (m). The thermal conductivity $k$ is estimated by the seven schemes mentioned above, and other soil parameter calculations are fixed as their original schemes, with the effects of SOM and gravels also included as implemented in thermal conductivity schemes (see details in *Dai et al.* [2014]).

The comparison of CoLM incorporating different soil thermal conductivity schemes is conducted against in-situ observations from five sites. Four of the sites (US-FPe, IT-Ro1, US-Bo1 and AU-How) are from FLUXNET network which provides long-term observed ground surface temperature and energy components in ground surface energy balance [*Baldocchi et al.*, 2001], and the other one is the Nagqu site from the Qinghai - Tibet Plateau where the observed data include not only ground surface variables but also soil temperature and moisture profiles measured at depths of 5, 10, 20, 40, 80 and 160 cm [*Pan et al.*, 2017]. These sites are selected mostly based on their annual amounts of precipitation, so that they can cover a wide range of climate conditions from arid (US-FPe) to moist (AU-How). The basic information of these sites and the duration of simulation over each are listed in Table 1. The first several years in each CoLM run are used for spin-up to make the fields in deep soil layers reach equilibrium, and the simulated results for the last year are used for analyses. The Nagqu site provides both atmospheric forcings and soil property information for CoLM runs, while the sites from FLUXNET only provide atmospheric forcings and the corresponding basic soil information are from the global soil datasets GSDE [*Shangguan et al.*, 2014], which provides the fractions of mineral soils, organic matter and bulk density, and SoilGrids [*Hengl et al.*, 2017], which provides the fractions of gravels. The two datasets have the kilometer and sub-kilometer spatial resolutions, and thus the soil information can be accurately extracted for use at the simulating sites.

2.2.3 The uncertainties of CoLM simulations from in-situ to global scale introduced by thermal conductivity formulation

The above two steps of evaluations can examine the relative performances of soil thermal conductivity schemes in both soil science and land modelling applications, and based on that, a superior scheme can be identified for land modelling use. Next, in order to tell to what extent soil thermal conductivity formulation can affect land surface modelling, the inter-scheme differences of the simulated results by CoLM are analyzed to quantify the induced uncertainties by different thermal conductivity schemes. First, we mainly concentrate on the situations over two sites, Nagqu and AU-How, which represent dry and wet conditions, respectively. The relationship between the inter-scheme differences of the simulated soil thermal conductivity and soil temperature are particularly investigated. The corresponding soil water profiles, ice content and precipitation patterns are analyzed as well, given the strong control of soil water content and status on soil thermal conductivity. Note that we choose the site Nagqu to represent dry conditions instead of the drier site US-FPe for consistency with the evaluation analyses in which only the soil temperature profiles at Nagqu are provided.

Then, we implement a global comparison between CoLM performances with the identified superior scheme and the mean of the other schemes to see how the soil thermal conductivity affects land surface modelling on the global scale. The two simulations are performed for three years (2000-2002), with the first two years used for spin-up and the last year used to form annual averages for simulated variables. Similar to the in-situ runs over the FLUXNET sites, the basic soil information for global runs are provided by the GSDE and SoilGrids datasets, which have a spatial resolution of 1 km. In order to eliminate uncertainties introduced by upscaling processes, our simulations follow the super-high spatial resolution of the soil property datasets, and hence only two global cases instead of seven cases with individual thermal conductivity scheme are performed here to avoid too huge amounts of calculations and data storage. The atmospheric forcings are from the version 7 of Climate Research Unit–National Centers for Environmental Prediction (CRUNCEP) data with QIAN-NCEP ocean fill (see https://www.earthsystemgrid.org/dataset/ucar.cgd.ccsm4.CRUNCEP.v4.html for download information), which are interpolated into the spatial resolution of the soil property datasets. In the analyses, the global patterns of the differences of simulated soil temperature and moisture at multiple layers, ground surface fluxes and snow cover are all presented. The results can highlight the necessity of the evaluation and development of soil thermal conductivity schemes with respect to land modelling applications.

2.2.4 Statistical metrics

The major metrics for scheme calibration are the bias, root-mean-squared error (RMSE) and relative error ($\sigma$) defined as:

$$Bias = \frac{1}{N}\sum_{1}^{N}\left(k_{pre} - k_{mean}\right) \qquad (25)$$

$$RMSE = \sqrt{\frac{1}{N}\sum_{1}^{N}\left(k_{pre} - k_{mean}\right)^{2}} \qquad (26)$$

$$\sigma = \frac{RMSE - median(RMSE)}{median(RMSE)} \qquad (27)$$

where $k_{pre}$ and $k_{mean}$ are the predicted and measured soil thermal conductivities, $N$ is the total number of sample data, and median(*RMSE*) is the median value of RMSE of all the schemes. The relative error depicts the relative performance of each thermal conductivity scheme, indicating how much better or worse than the median level of scheme predictions [*Gleckler et al.*, 2008]. In CoLM simulations, the bias and RMSE are used to evaluate the model performances. We limit our analyses to daily means of model output to minimize the uncertainties of observations due to measurement. The inter-scheme differences for land surface modelling are represented by the standard deviations of the simulated results with different thermal conductivity schemes.

## 3. Results and Discussion

### 3.1 *Evaluations of soil thermal conductivity schemes*

Figure 2 presents the comparisons of the seven soil thermal conductivity schemes using the experimental measurements from the adopted soil samples in this study. On visual inspection, all the schemes are able to produce acceptable predictions. The dry thermal conductivities are quite small, and thus their predictions are closest to the measurements. As the soils become saturated, the magnitudes of thermal conductivity increase and their predictive deviations are increasingly large. Statistics analyses show that the best estimates of thermal conductivity are given by the *Balland and Arp* [2005] scheme, with the overall bias of 0.01 W/m/K and RMSE of 0.28 W/m/K. The *Johansen* [1975] scheme and *Côté and Konrad* [2005] scheme also provide satisfactory estimates, whose biases are 0.01 and 0.03 W/m/K and RMSE are 0.30 and 0.32 W/m/K, respectively. The two mechanistic schemes, *Tarnawski and Leong* [2012] and *De Vries* [1963], stay in the median level in which the predictions underestimate the measurements in above half-saturated soil conditions. The *Lu et al.* [2007] scheme and *Farouki* [1981] scheme significantly overestimate the measurements at most soil saturation levels, which may partly result from the changes of the two schemes from *Johansen* [1975] with respect to the formulation of dry (i.e. Equation 16) and saturated (i.e. Equation 9-10) thermal conductivity.

To further ascertain the relative performance of the seven schemes in different soil conditions, the relative error based on the RMSE of each scheme is calculated at all the measured saturation levels (Figure 3). The results show that the *Côté and Konrad* [2005] scheme produces the closest estimates for the dry soils ($S_r = 0$), whose RMSE is at least 30% smaller than the median error, while three of the schemes, *Lu et al.* [2007], *Tarnawski and Leong* [2012] and *De Vries* [1963], deviate most from the measurements (over 30% larger RMSE than the median error). The *Balland and Arp* [2005] scheme and *Tarnawski and Leong* [2012] scheme provide the best fit over low-saturation levels ($S_r = 0.1, 0.25$), while the *Lu et al.* [2007] scheme still shows the least accuracy. The relatively poor performance of *Lu et al.* [2007] is probably due to its simple linear fitting for the dry thermal conductivity without consideration of soil type effects [*Zhang and Wang*, 2017]. At saturations from 0.5 to 1.0, the top estimates of thermal conductivity are made by the *Balland and Arp* [2005] scheme. The *Farouki* [1981] scheme gives the relatively poorest predictions, which is consistent with its systematic overestimations as shown in Figure 2, likely resulting from its biased parameterization of the saturated thermal conductivity. Overall, the relative errors demonstrate that the *Balland and Arp* [2005] scheme consistently ranks on or near the top among all the schemes over various soil saturation levels, which can be related to the continuous description of thermal conductivity variations by its $K_e$ - $S_r$ relationship. This is also pointed out in the assessment analyses of *Zhang and Wang* [2017].

Next, we evaluate the thermal conductivity schemes with respect to their land modelling applications based on their corresponding CoLM performances over the five observational sites. Figure 4 presents the simulated and observed soil temperature variations over the Nagqu site. It can be seen that the closest soil temperature simulations in the depth of 5 cm and 40 cm are both produced by CoLM with the *Balland and Arp* [2005] scheme, whose RMSE are 1.93 and 1.03 K, respectively. Consistent results can be obtained at other measured depths (not shown here for simplicity), indicating that the *Balland and Arp* [2005] scheme performs best across the whole soil column from top to bottom. Comparisons of the simulated ground heat flux also reveal the best performance for the *Balland and Arp* [2005] scheme (Table 2), with which an average of 3 W/m$^2$ lowered RMSE is obtained than that calculated with the other schemes. Except ground heat flux which depends on the soil thermal conductivity at surface, the other energy components in ground surface energy balance in the simulations seem to be insensitive to thermal conductivity calculations (Table 2). This is not surprising as the simulations of turbulent and radiative fluxes are mostly determined by the descriptions of land surface properties and parameterizations with respect to land-atmosphere interactions [*Dai et al.*, 2003]. Table 2 also lists the error statistics of the simulated variables with each thermal conductivity scheme over the other four sites which cover climatic conditions from dry to wet. The results show that the smallest RMSE of ground surface temperature and ground heat flux over all the sites are consistently obtained from the simulations with the *Balland and Arp* [2005] scheme, which suggests that the *Balland and Arp* [2005] scheme is more suitable for land surface modelling than other schemes over various climates.

In summary, the *Balland and Arp* [2005] scheme has exhibited the best performance among the selected thermal conductivity schemes with respect to both scheme estimation and land process simulation, and thus can be chosen as a superior scheme for land modelling use. Although the inter-scheme differences of the mean errors of CoLM-simulated results seem to be not that remarkable at ground surface, those produced at deep layers can make significant responses to thermal conductivity variations through the changes of soil heat conduction and the associated freeze-thaw cycles, which has been partly reflected by the larger spread of the simulated soil temperature at the depth of 40 cm than that at 5 cm over the Nagqu site as shown in Figure 4. This will be further illustrated by analyzing the uncertainties of CoLM simulations introduced by thermal conductivity formulation in the next section.

3.2 *Quantifications of uncertainties in land surface modelling introduced by thermal conductivity formulation*

The *Balland and Arp* [2005] scheme has been shown to perform better than the other six schemes in land surface processes modelling. Here we quantify the uncertainties of CoLM simulations introduced by thermal conductivity formulation to investigate the sensitivity of CoLM to thermal conductivity variations. First, the situations over the two sites, Nagqu and AU-How, are analyzed. The former site is characterized as dry and partly frozen conditions, and the latter is characterized as wet and non-frozen conditions.

Figure 5a presents the inter-scheme differences of the simulated soil thermal conductivity profiles over the Nagqu site. In warming seasons, the most significant differences (on the order of 0.5 W/m/K) appear in shallow layers when soil moisture becomes relatively high following precipitation variations. Their excellent correspondence shown in Figure 5a, 5b and 5d is expected as high soil saturations have prominent effects on increasing the bulk thermal conductivity, and thus the inter-scheme differences of thermal conductivity become noticeable. The simulated soil temperature at shallow layers does not respond to the inter-scheme variations of soil thermal conductivity (Figure 5e), since the temperature near the surface remains controlled by ground surface energy balance. However, due to the changes of soil heat flux transport from top to bottom induced by thermal conductivity variations near the surface, the simulated soil temperature exhibits an increasingly large spread towards deep layers (Figure 5e), with the magnitude up to 2.5 K at the bottom. This reflects a vertically accumulative effect of thermal conductivity variations on soil temperature simulations. In cold seasons, although the changes of thermal conductivity across the frozen layers are significant (Figure 5a, 5c) due to the greater thermal conductivity of ice (2.29 W/m/K) than liquid water (0.57 W/m/K), the simulated soil temperature is not sensitive to thermal conductivity variations within the frozen period (Figure 5e). This may be attributable to surface cooling for which few energy can be obtained by soils to transfer downwards and thus soil heat fluxes are mostly unchanged. It is noticeable that the significant inter-scheme differences of the simulated soil temperature appear around the dates of fall ice freeze-up and spring ice break-up across almost all soil layers (Figure 5e), which corresponds well to the patterns of differences of the simulated soil ice content (Figure 5f). Very likely the changes of soil heat conduction due to different thermal conductivity calculations can lead to several days shift of ice freeze-up and break-up dates, during which the differences of the simulated latent heat release or absorption due to phase change of soil water can trigger large inter-scheme variations of soil temperature simulations.

Unlike the clear responses to the variations of soil thermal conductivity over the Nagqu site, the simulated soil temperature at AU-How seems to be independent of the thermal conductivity calculations during the entire simulating periods (Figure 6d). The inter-scheme differences of soil thermal conductivity are notable across all the soil layers due to high soil moisture levels with large amounts of precipitation (Figure 6a, 6b, 6c), but the high soil moisture can also lead to a large soil heat capacity and thus increase the soil heat inertia. Therefore, the simulated soil temperature displays little sensitivity to thermal conductivity variations. Overall, under relatively dry conditions, the simulated soil temperature is significantly affected by thermal conductivity variations especially at deep layers due to changes of soil heat conduction and the associated freeze-thaw cycles, while in wet conditions, the large soil heat inertia can make the simulations stable among different thermal conductivity schemes.

Next, we conduct a global comparison between CoLM performances with the identified superior scheme, the *Balland and Arp* [2005] scheme, and the mean of the other schemes to see how the soil thermal conductivity affects land surface modelling on the global scale. Figure 7 shows the global distribution of the simulated soil temperature differences over multiple layers. Consistent with the in-situ simulated results, the most significant differences appear over arid and semi-arid regions in mid- and high-latitudes, and the pattern becomes increasingly noticeable towards deep layers with the magnitude of differences more than 1 K over northern frozen areas. Examinations for the differences of ground surface energy components reveal that the ground heat flux is significantly affected by thermal conductivity variations (Figure 8a), with about 2 W/m$^2$ more energy transported into soils which results in the overall warming across soil layers shown in Figure 7. The differences of ground heat flux are mostly balanced by those from sensible heat flux (Figure 8b), while the differences of latent heat flux and net radiative flux are typically small. Considering the energy redistribution can influence the phase change of snow cover and soil water, here we also examine the differences of related hydrological variables. As results of the larger ground heat flux simulated by the *Balland and Arp* [2005] scheme, the snow cover and soil moisture at shallow layers tend to decrease over high latitudes due to the melt of snow and ice (Figure 8e, 8f), and the consequent more infiltrated water tends to increase the soil water content with depth (Figure 8g, 8h). Given the above analyses, a slight positive feedback is likely formed as follows: the increased soil moisture can lead to an increase of soil thermal conductivity at deep layers, and as a response, the enhanced ground heat flux can further melt ice and snow at surface layers, and hence the more infiltrated water can keep the soil moisture in high levels at deep layers, which continuously affects soil thermal conductivity. This can also partly explain the stronger responses of the simulated soil temperature at deep layers as shown in Figure 7.

    The above analyses illustrate that the soil thermal conductivity can significantly affect land surface modelling over relatively dry and cold regions, where the simulated results are sensitive to thermal conductivity variations especially for the soil temperature at deep layers. The changes of soil energy and water transport and the associated phase change of water are all responses to thermal conductivity variations. Due to too huge amounts of calculations and data storage, we do not investigate the

effects of soil thermal conductivity on climate using coupled earth system models. In fact, the climate over boreal dry and cold regions has been shown to be sensitive to external forcings due to its stable conditions with a small atmospheric effective heat capacity [e.g. *Davy and Esau*, 2014a; 2014b], and thus even a slight difference in simulated ground surface temperature and heat flux transported into the atmosphere can lead to a large difference in predicted atmospheric temperature, boundary layer height, cloud cover fractions and many other related circulation systems [e.g. *Lawrence and Slater*, 2008; *Wei et al.*, 2014]. Moreover, the changes of snow cover and soil freeze-thaw cycles over high latitudes can alter the atmospheric moisture budget through regulating surface evaporation and water vapor advection, and thus the global hydrological cycles and atmospheric circulation are affected [e.g. *Callaghan et al.*, 2012]. These results suggest a critical role of soil thermal conductivity in climate systems, implying that further development of thermal conductivity schemes can even alter the accuracy of weather and climate models by correcting the simulations of land-atmospheric interactions.

Our results present the best performance for the *Balland and Arp* [2005] scheme, which infers that the *Farouki* [1981] scheme, the default scheme to estimate soil thermal conductivity in many land surface models, is not a preferable choice for land modelling use. However, one of the major advantages of the *Farouki* [1981] scheme is its global data availability, which can be obtained by several soil datasets, although we have tested the relatively low accuracy of this scheme in middle and high saturations. In fact, to keep the high accuracy of the *Balland and Arp* [2005] scheme for global land surface modelling, a global dataset for the parameters in this scheme needs to be built. Based on the fact that the fast enrichment of soil property datasets has been undergoing, it is possible to fit the parameters of the *Balland and Arp* [2005] scheme with the collected soil information at every global grid in the future.

4. **Conclusion and future prospects**

In this work, seven highly recommended soil thermal conductivity schemes with the modification of containing soil organic matter and gravel effects are evaluated for their applicability in land surface modeling. The accuracy of the scheme estimation and land process simulation are both considered in the evaluation. By direct comparison to laboratory measured thermal conductivities over a range of soil saturation levels and types, and by comparing land modelling performances incorporating different thermal conductivity schemes to observed in-situ land state variables using the CoLM, the *Balland and Arp* [2005] scheme is found to consistently perform best among all the schemes, and thus can be recommended as a superior scheme for land surface modeling use.

In order to tell to what extent land surface modeling relies on soil thermal conductivity schemes, the uncertainties of CoLM simulations introduced by thermal conductivity formulation are investigated. The results demonstrate that, over relatively dry regions, the inter-scheme variations of soil thermal conductivity can significantly lead to differences of simulated soil temperature, especially at deep layers due to changes of downward soil heat conduction and the associated freeze-thaw cycles. However, the thermal conductivity seems to have fewer effects on simulations over wet regions, likely due to the high soil heat capacity induced by high soil moisture levels, which increases the heat inertia in soil thermodynamics. Global comparisons between the CoLM performances with the *Balland and Arp* [2005] scheme and the mean of the other schemes also infer that the differences of thermal conductivity can introduce significant changes of soil temperature over arid and semi-arid regions in mid- and high-latitudes, particularly in deep layers, and the associated changes of ground heat flux, snow cover and soil moisture appear in correspondence. These results display the role of soil thermal conductivity in land surface modeling, and suggest the importance of the evaluation and further development of thermal conductivity schemes with respect to land modelling applications.

We realize that the performances of soil thermal conductivity schemes can be significantly affected by the estimations of soil hydraulic parameters due to the strong coupling in simulating soil energy and water transport. Soil hydraulic parameters are always calculated as the ensemble means of the predictions from multiple pedo-transfer functions with the basic soil information [e.g. *Dai et al.*, 2013]. Several regional measurements for soil water retention have shown that the estimations using this method are reasonable in land surface modelling [e.g. *Zhang et al.*, 2016]. However, this method does not contain the effects of soil thermal parameters either. Future works will focus on the effects of soil thermal and hydraulic parameters on each other, and conduct joint evaluations and development for their calculation schemes using more available observations. This can also lay the foundation for upscaling of these parameters when coupling with global climate models.

**Acknowledgments**

This research was supported by the Natural Science Foundation of China (under grant 41805072 and 41730962), and National Key R&D Program of China (grant no. 2017YFA0604300 and 2016YFB0200801). We are grateful to Prof. Yingping Wang and Prof. Xinzhong Liang for their helpful suggestions. We also thank Dr. Yufei Xin for providing observed datasets over the Nagqu site.

Table 1. Basic information of the sites used for evaluations of CoLM performances

| Site | Country | Location | Land cover type (USGS) | Precipitation (mm/year) | Duration of simulation |
|---|---|---|---|---|---|
| US-FPe | USA | 48.3°N, 105.1°W | Grassland | 335 | 2000.01-2006.12 |
| Nagqu | China | 31.37°N, 91.9°E | Grassland | 420 | 2011.07-2015.07 |
| IT-Ro1 | Italy | 42.4°N, 11.9°E | Deciduous broadleaf forest | 764 | 2002.01-2006.12 |
| US-Bo1 | USA | 40.0°N, 88.3°W | Cropland and Pasture | 1066 | 1997.01-2006.12 |
| AU-How | Australia | 12.5°S, 131.2°E | Savanna | 1449 | 2002.01-2005.12 |

**Table 2**. Root mean square errors of the CoLM-simulated ground surface temperature and ground surface energy components with the seven soil thermal conductivity schemes over the five observational sites.

|        | Farouki [1981] | Johansen [1975] | Côté and Konrad [2005] | Balland and Arp [2005] | Lu et al. [2007] | Tarnawski and Leong [2012] | De Vries [1963] |
|--------|---------|---------|---------|---------|---------|---------|---------|
|        |         |         |         | Naqu    |         |         |         |
| $T_s$  | **2.06** | **2.01** | **2.04** | **1.93** | **2.06** | **2.19** | **2.14** |
| $G$    | **29.83** | **30.16** | **30.18** | **27.63** | **30.19** | **31.06** | **29.91** |
| $LH$   | 21.99 | 21.36 | 22.15 | 21.22 | 22.19 | 21.45 | 21.97 |
| $SH$   | 35.89 | 37.25 | 36.1  | 37.38 | 35.96 | 37.29 | 36.53 |
| $R_{net}$ | 13.59 | 13.52 | 13.42 | 13.58 | 13.4 | 13.62 | 13.5 |
|        |         |         |         | US-FPe  |         |         |         |
| $T_s$  | **3.51** | **3.57** | **3.52** | **3.4** | **3.5** | **3.53** | **3.52** |
| $G$    | **26.82** | **22.77** | **22.48** | **19.88** | **23.8** | **23.26** | **23.34** |
| $LH$   | 23.43 | 23.48 | 23.48 | 23.49 | 23.46 | 23.48 | 23.49 |
| $SH$   | 29.53 | 25.32 | 26.14 | 25.42 | 27.03 | 25.91 | 26.91 |
| $R_{net}$ | 22.28 | 22.43 | 22.43 | 22.4 | 22.32 | 22.36 | 22.35 |
|        |         |         |         | IT-Ro1  |         |         |         |
| $T_s$  | **2.68** | **2.66** | **2.68** | **2.6** | **2.68** | **2.66** | **2.69** |
| $G$    | **15.24** | **14.05** | **14.23** | **11.03** | **13.73** | **14.21** | **13** |
| $LH$   | 31.56 | 32.47 | 31.47 | 32.41 | 31.44 | 32.3 | 31.4 |
| $SH$   | 40.68 | 39.8 | 40.88 | 39.76 | 40.79 | 39.69 | 40.76 |
| $R_{net}$ | 29.11 | 29.12 | 29.12 | 29.11 | 29.13 | 29.08 | 29.1 |
|        |         |         |         | US-Bo1  |         |         |         |
| $T_s$  | **2.27** | **2.21** | **2.21** | **2.18** | **2.23** | **2.22** | **2.23** |
| $G$    | **19.6** | **15.77** | **15.5** | **13.7** | **16.62** | **16.09** | **16.14** |
| $LH$   | 25.44 | 26.14 | 25.14 | 26.16 | 25.05 | 26.03 | 25.02 |
| $SH$   | 26.59 | 24.38 | 25.35 | 24.34 | 25.61 | 24.83 | 25.68 |
| $R_{net}$ | 21.47 | 21.39 | 21.38 | 21.4 | 21.42 | 21.41 | 21.4 |
|        |         |         |         | AU-How  |         |         |         |
| $T_s$  | **1.74** | **1.77** | **1.79** | **1.72** | **1.77** | **1.74** | **1.77** |
| $G$    | **9.92** | **8.42** | **8.98** | **6.67** | **8.84** | **8.73** | **8.15** |
| $LH$   | 37.97 | 36.98 | 37.93 | 36.9 | 37.95 | 37.01 | 37.94 |
| $SH$   | 26.09 | 27.68 | 26.4 | 27.45 | 26.47 | 28 | 26.67 |
| $R_{net}$ | 21.51 | 21.32 | 21.52 | 21.43 | 21.46 | 21.22 | 21.37 |

$T_s$ represents ground surface temperature, $G$ represents ground heat flux, $LH$ and $SH$ represent latent and sensible heat flux, and $R_{net}$ represents net radiative flux.

**Figure Captions**

**Figure 1.** Soil thermal conductivities of the 40 Canadian soils as a function of saturation levels. The color of each line denotes the porosity of each soil sample.

**Figure 2.** Comparisons of the predicted soil thermal conductivities by the seven schemes with laboratory measurements from the adopted soil samples over multiple saturation levels. The straight line is the 1:1 line.

**Figure 3.** Relative errors of the seven soil thermal conductivity schemes in terms of 6 saturation levels. The error measure, treating each saturation level separately, is calculated by normalizing the root mean square error (RMSE) of each scheme by the median RMSE of all the schemes, with blue shading indicating performance being better, and red shading worse, than the median level of scheme predictions. $k_{dry}$, $k_{S_r=0.1}$, $k_{S_r=0.25}$, $k_{S_r=0.5}$, $k_{S_r=0.7}$, $k_{sat}$ represent the soil thermal conductivity at the saturation levels 0, 0.1, 0.25, 0.5, 0.7, 1. Dat1 means that the relative error is measured based on the 40 Canadian soil samples, and Dat2 represents the corresponding values from the 21 other samples.

**Figure 4.** CoLM-simulated soil temperature with the seven thermal conductivity schemes over the Nagqu site at soil depth of (a) 0.05m and (b) 0.4m.

**Figure 5.** Vertical profiles of simulated (a) standard deviation of soil thermal conductivity, (b) mean soil moisture, (c) mean volumetric soil ice content, (e) standard deviation of soil temperature and (f) standard deviation of volumetric soil ice content with different soil thermal conductivity schemes over the Nagqu site with (d) precipitation variations given as well.

**Figure 6.** Vertical profiles of simulated (a) standard deviation of soil thermal conductivity, (b) mean soil moisture and (d) standard deviation of soil temperature with different soil thermal conductivity schemes over the AU-How site with (c) precipitation variations given as well.

**Figure 7.** Global patterns of differences of CoLM-simulated soil temperature with the

*Balland and Arp* [2005] scheme and the mean of the other schemes at (a) the top three layers, (b) the fifth layer and (c) the seventh layer.

**Figure 8.** Global patterns of differences of CoLM-simulated (a) ground heat flux, (b) sensible heat flux, (c) latent heat flux, (d) net radiative flux, (e) snow water equivalent and soil moisture at (f) the top three layers, (g) the fifth layer and (h) the seventh layer with the *Balland and Arp* [2005] scheme and the mean of the other schemes.

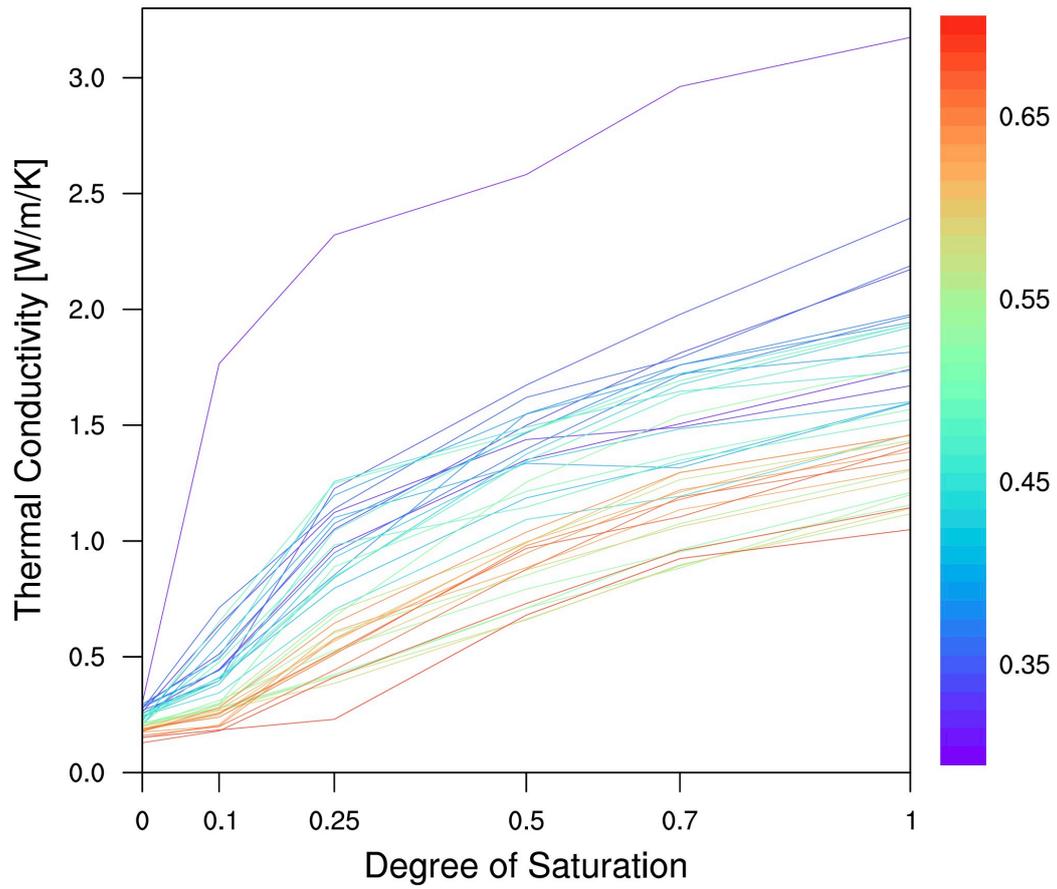

**Figure 1.** Soil thermal conductivities of the 40 Canadian soils as a function of saturation levels. The color of each line denotes the porosity of each soil sample.

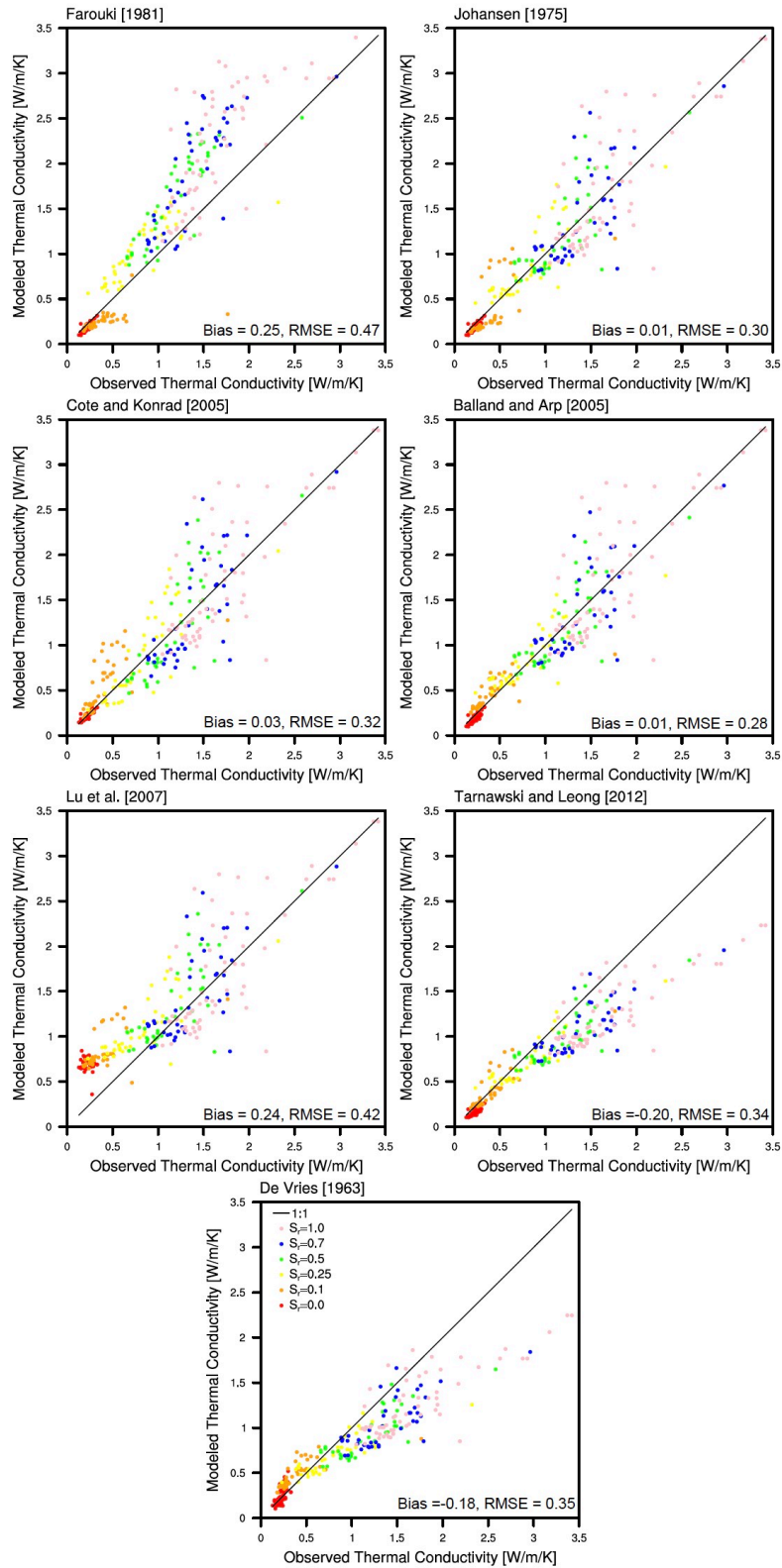

**Figure 2.** Comparisons of the predicted soil thermal conductivities by the seven schemes with laboratory measurements from the adopted soil samples over multiple saturation levels. The straight line is the 1:1 line.

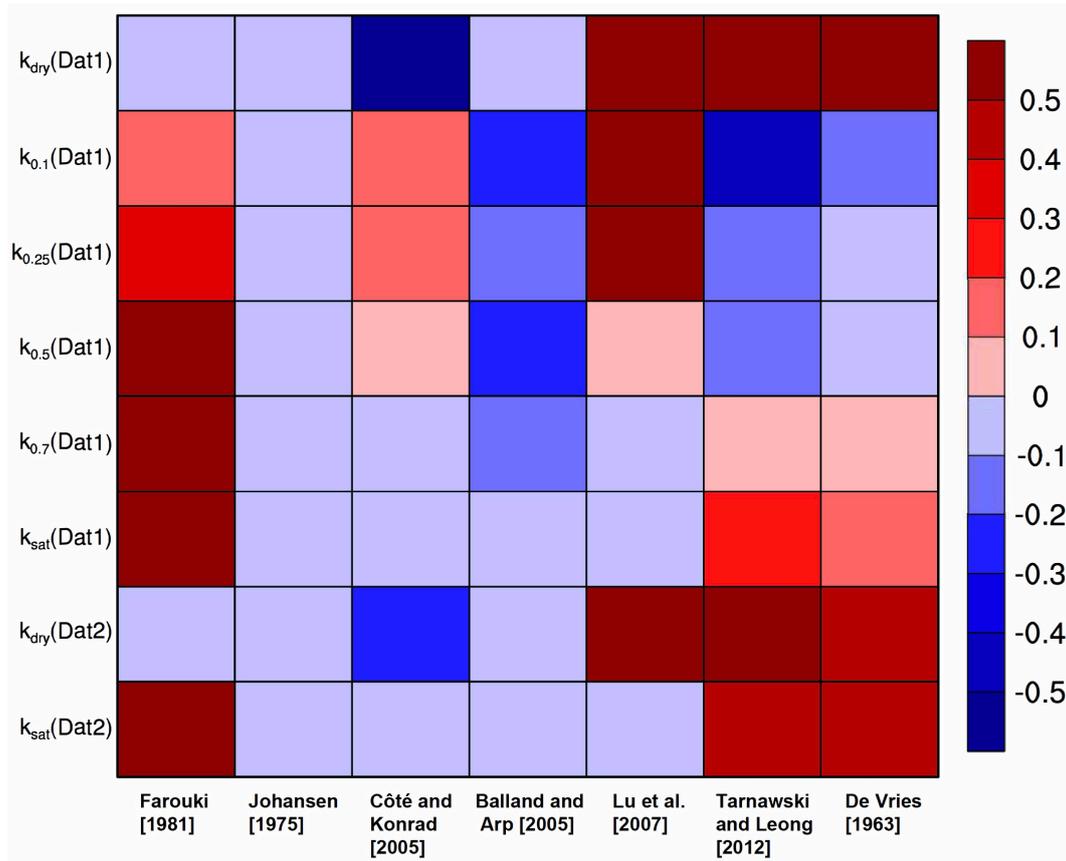

**Figure 3.** Relative errors of the seven soil thermal conductivity schemes in terms of 6 saturation levels. The error measure, treating each saturation level separately, is calculated by normalizing the root mean square error (RMSE) of each scheme by the median RMSE of all the schemes, with blue shading indicating performance being better, and red shading worse, than the median level of scheme predictions. $k_{dry}$, $k_{S_r=0.1}$, $k_{S_r=0.25}$, $k_{S_r=0.5}$, $k_{S_r=0.7}$, $k_{sat}$ represent the soil thermal conductivity at the saturation levels 0, 0.1, 0.25, 0.5, 0.7, 1. Dat1 means that the relative error is measured based on the 40 Canadian soil samples, and Dat2 represents the corresponding values from the 21 other samples.

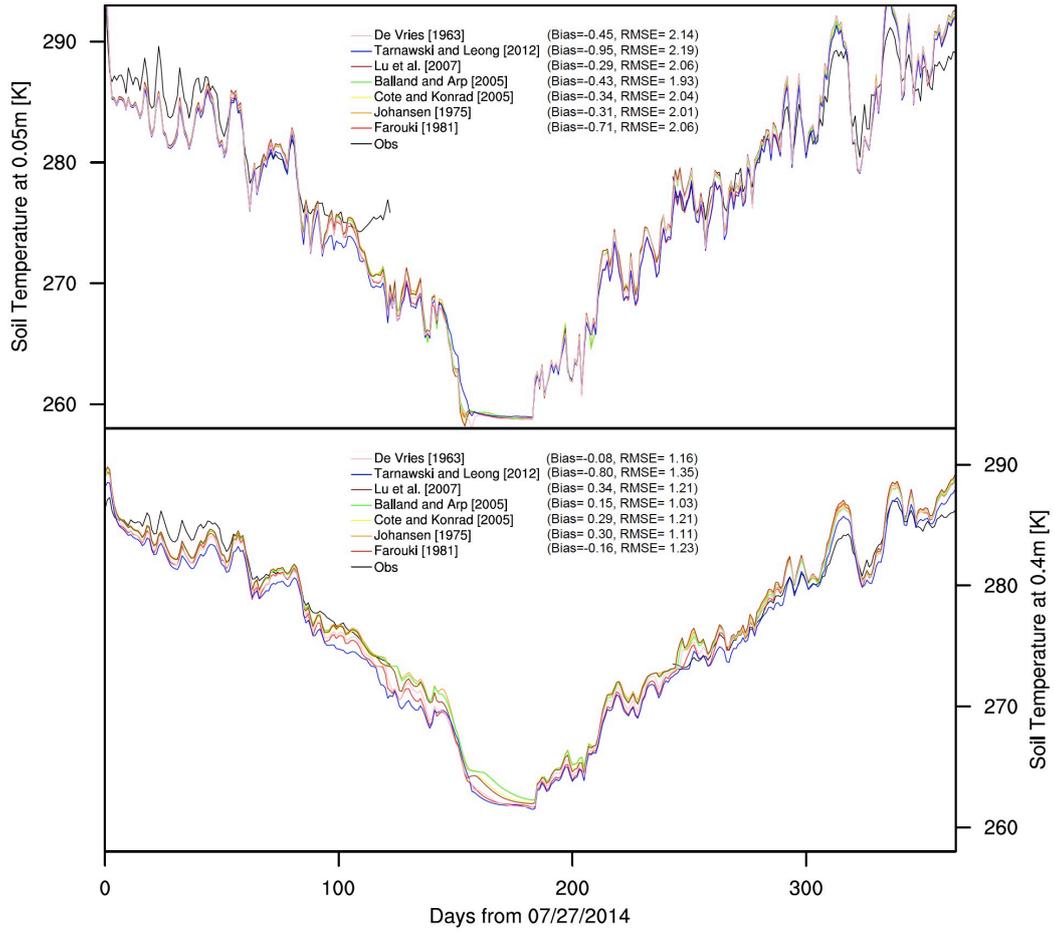

**Figure 4.** CoLM-simulated soil temperature with the seven thermal conductivity schemes over the Nagqu site at soil depth of (a) 0.05m and (b) 0.4m.

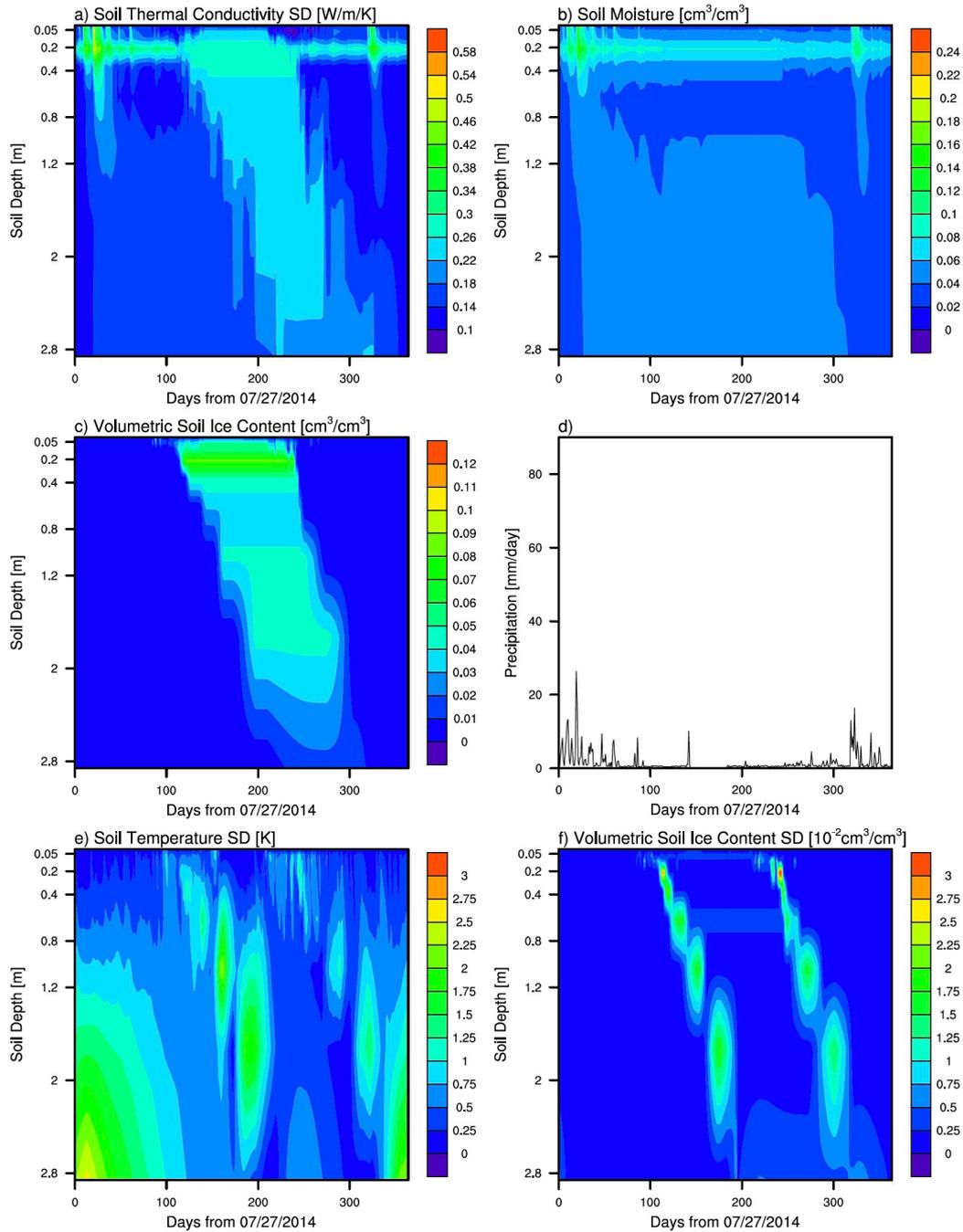

**Figure 5.** Vertical profiles of simulated (a) standard deviation of soil thermal conductivity, (b) mean soil moisture, (c) mean volumetric soil ice content, (e) standard deviation of soil temperature and (f) standard deviation of volumetric soil ice content with different soil thermal conductivity schemes over the Nagqu site with (d) precipitation variations given as well.

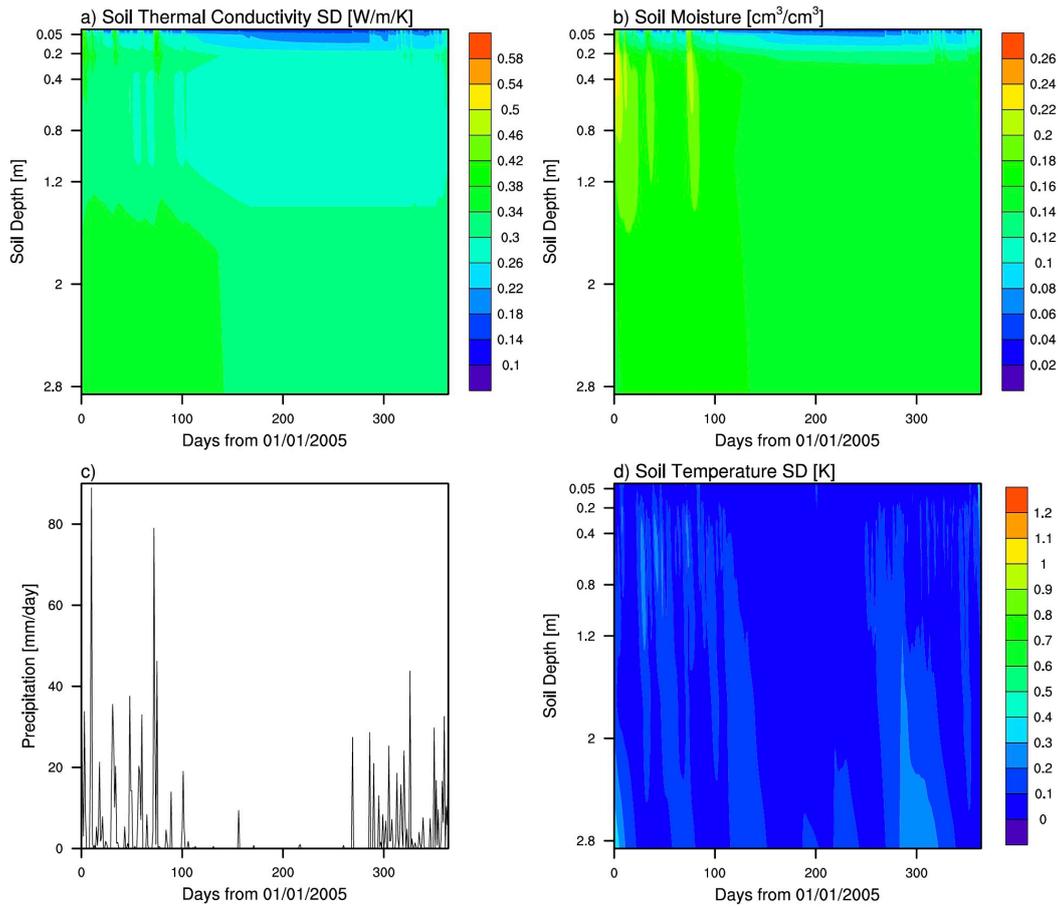

**Figure 6.** Vertical profiles of simulated (a) standard deviation of soil thermal conductivity, (b) mean soil moisture and (d) standard deviation of soil temperature with different soil thermal conductivity schemes over the AU-How site with (c) precipitation variations given as well.

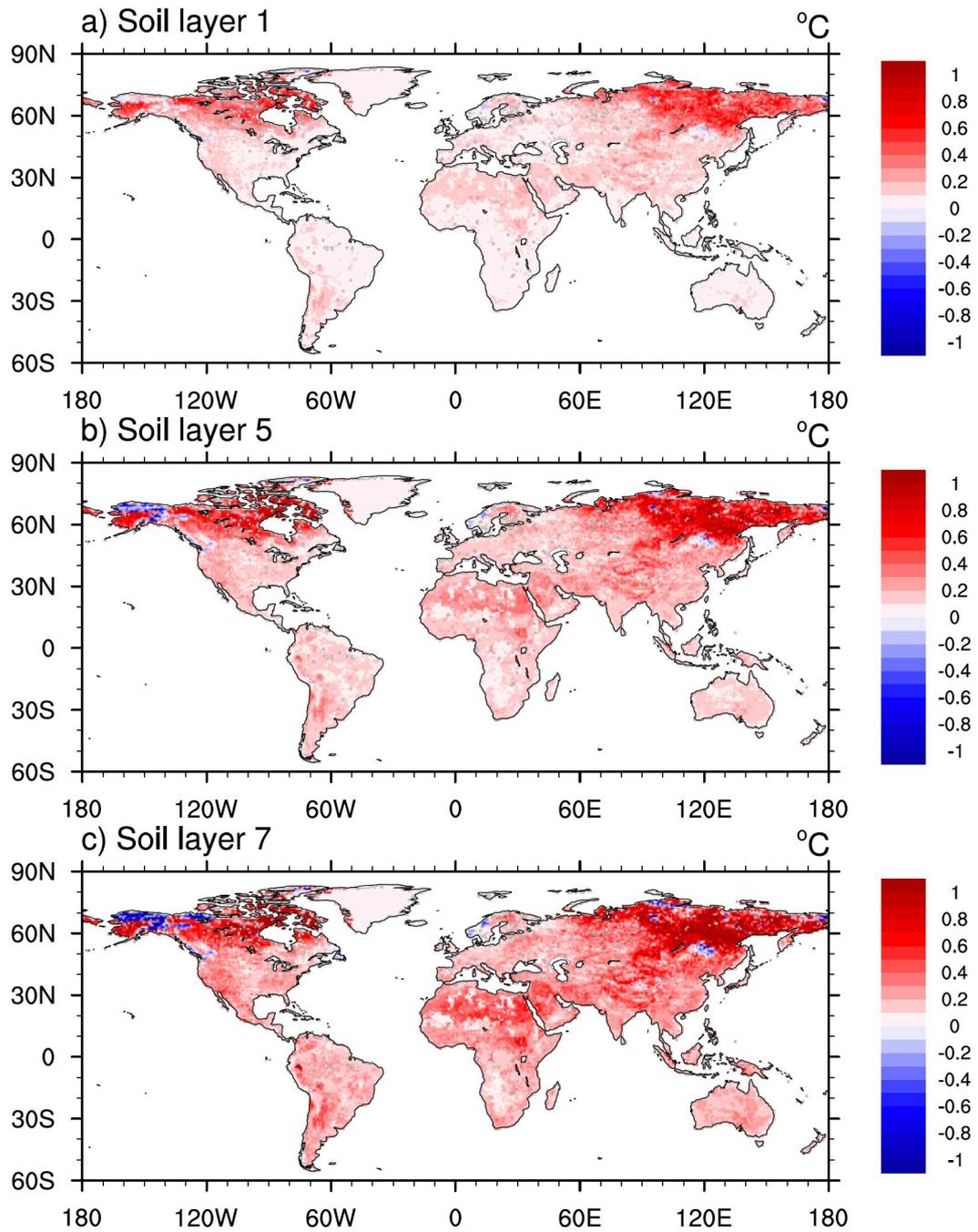

**Figure 7.** Global patterns of differences of CoLM-simulated soil temperature with the *Balland and Arp* [2005] scheme and the mean of the other schemes at (a) the top three layers, (b) the fifth layer and (c) the seventh layer.

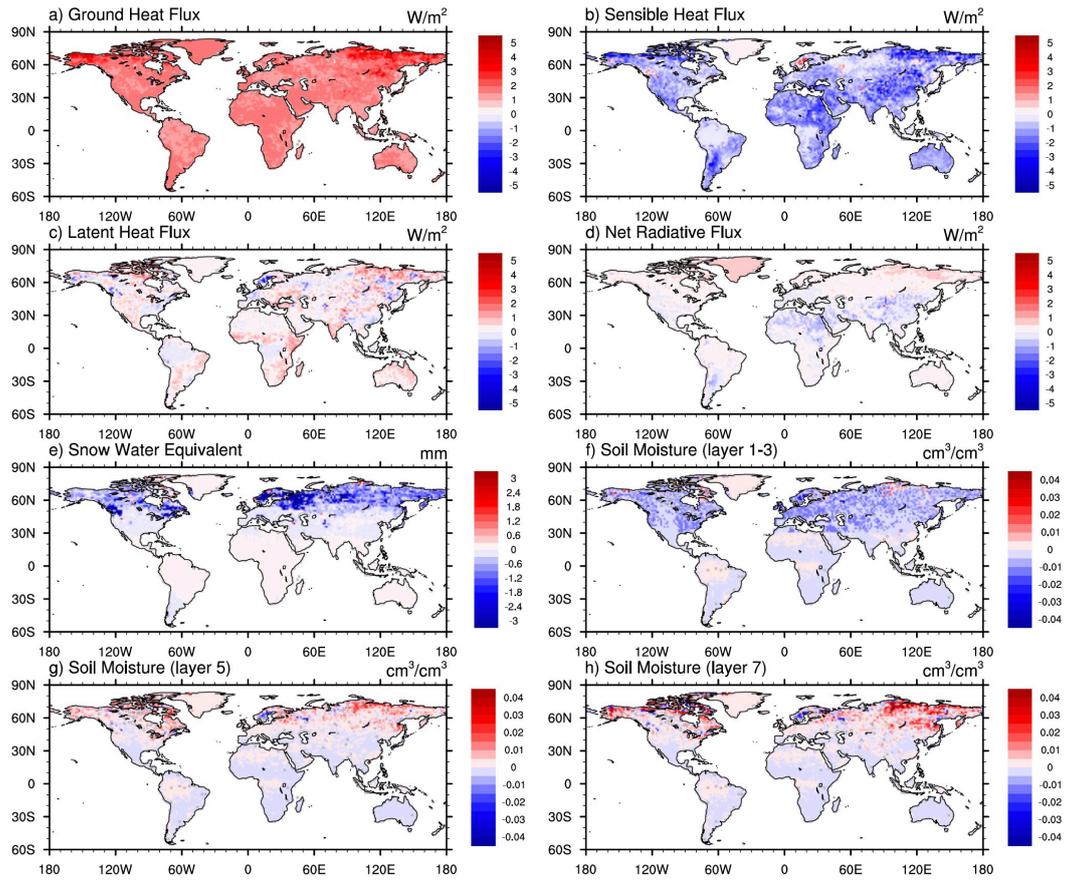

**Figure 8.** Global patterns of differences of CoLM-simulated (a) ground heat flux, (b) sensible heat flux, (c) latent heat flux, (d) net radiative flux, (e) snow water equivalent and soil moisture at (f) the top three layers, (g) the fifth layer and (h) the seventh layer with the *Balland and Arp* [2005] scheme and the mean of the other schemes.